\documentstyle[11pt]{l-aa}

\begin{document}

\title{GRB afterglows: from ultra-relativistic to non-relativistic phase}

\author{ Y.F.~Huang\inst{1}, Z.G.~Dai\inst{1}, and T.~Lu\inst{2,1} }

\offprints{ T.~Lu (E-mail: tlu@nju.edu.cn) }

\institute{ Department of Astronomy, Nanjing University, Nanjing 210093, 
	    P.R. China
\and        CCAST, Beijing 100080, P.R. China
	  }

\thesaurus{13.07.1, 02.19.1, 08.14.1, 09.10.1, 02.18.8}

\date{Received  ; accepted }

\maketitle
\markboth{Y.F. Huang et al.: Full evolution of GRB afterglows}{}  

\begin{abstract}

Postburst evolution of adiabatic fireballs that produce $\gamma$-ray 
bursts is studied. Emphasis has been put on the transition from the
highly relativistic phase to the non-relativistic phase, which, according 
to our calculation, should happen much 
earlier than previously expected. The theoretical light curves decline 
a little sharper at the non-relativistic stage than at early times, but 
still can fit the observations well. However, disagreements are obvious 
when $t \ge 100$ d, implying  a large initial energy $E_0$, a low 
interstellar density $n$, or the possible existence of a persistent 
energy source at the center of the fireball.  

\keywords{Gamma rays: bursts $-$ shock waves $-$ stars: neutron $-$ 
	  ISM: jets and outflows $-$ relativity}

\end{abstract}

\section{Introduction}

Since their discovery about thirty years ago (Klebesadel et al. 1973), 
$\gamma$-ray bursts (GRBs) have made one of the biggest mysteries in 
astrophysics (Fishman \& Meegan 1995). The Italian-Dutch BeppoSAX 
satellite opened up a new era in the field in early 1997. By the end of
May 1998, afterglows have been observed in X-rays from about a dozen 
events, in optical wavelengths in several cases (GRB 970228, 970508, 
971214, 980326, 980329, 980519, and possibly GRB 980425), and even 
in radio from GRB 970508, 980329, and GRB 980519. Very
recently, GRB 971214 was 
reported to have released an enormous energy of $3 \times 10^{53}$ erg 
in $\gamma$-rays alone (Kulkarni et al. 1998; Wijers 1998); and 
GRB 980425 seems to be associated with a supernova (Galama et al. 1998b), 
these facts have aroused researcher's more fever and may provide crucial 
clues to our understanding of GRBs. 

The fireball model has become the most popular and successful model of 
GRBs (M\'{e}sz\'{a}ros et al. 1994; Fenimore et al. 1996, and references 
therein). After producing the main GRB, the fireball will continue to 
expand as a thin shell into the interstellar medium (ISM), generating
an ultra-relativistic shock. Afterglows at longer wavelengths are 
produced by the shocked ISM (M\'{e}sz\'{a}ros \& Rees 1997;
Waxman 1997a,b; Tavani 1997; Sari 1997; Wijers et al. 1997; Huang 
et al. 1998; Dai \& Lu 1998a,b). It has been 
derived that for adiabatic expansion, $R \propto t^{1/4}$, 
$\gamma \propto t^{-3/8}$, where $R$ is the shock radius measured in 
the burster's static frame, $\gamma$ is the Lorentz factor of 
the shocked ISM measured by the observer and $t$ is the observed time. 
These scaling laws are valid only at the ultra-relativistic stage. 

The purpose of this {\em Letter} is to study numerically the full 
evolution of adiabatic fireballs from the ultra-relativistic phase 
to the non-relativistic phase. It is found that radiation during 
the mildly relativistic phase ($2 \leq \gamma \leq 5$) and the 
non-relativistic phase ($\gamma \leq 2$), which was obviously 
neglected in previous studies, is of great importance. 

\section{Previous studies} 

In case of adiabatic expansion, the shocked ISM's Lorentz factor
evolves based on (Sari 1997; Waxman 1997a,b; Tavani 1997): 
\begin{equation}
   \gamma \approx (200 - 400) E_{51}^{1/8} n_0^{-1/8} t_{\rm s}^{-3/8}, 
\end{equation}
where $E_0 = E_{51} \times 10^{51}$ erg is the original fireball 
energy, of which about a half is believed to have been released as 
$\gamma$-rays during the GRB phase, $n=n_0$ 1 cm$^{-3}$ is the 
number density of the unshocked ISM, and $t_{\rm s}$ is $t$ in units of 
second. $R(t)$ can be derived
from $\gamma^2 R^3 \approx E_0/(4 \pi n m_{\rm p} c^2)$, 
where $m_{\rm p}$ is the  
proton mass and $c$ the velocity of light. Flux density at observing 
\begin{figure}
\begin{picture}(100,160)
\put(0,0){\includegraphics{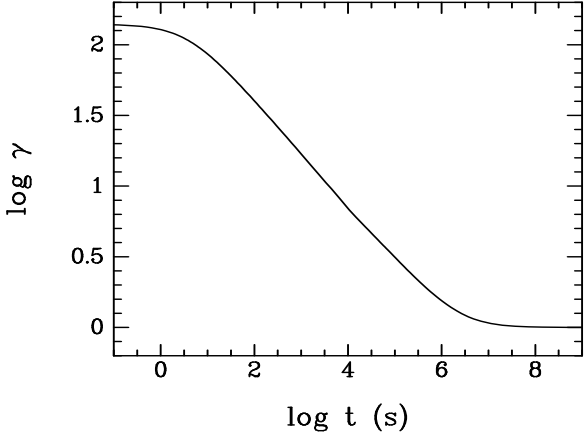}}
\end{picture}
\caption
{ Evolution of the fireball's Lorentz factor
}
\end{figure}
frequency $\nu$ then declines as $S_{\nu} \propto t^{3(1-p)/4}$, 
where $p$ is the index characterizing the power-law distribution of 
the shocked ISM 
electrons, $d n_{\rm e}'/d \gamma_{\rm e} \propto \gamma_{\rm e}^{-p}$. 
These expressions are valid only when $\gamma \gg 1$. In general, 
X-ray and optical afterglows were observed to follow power-law decays, 
and such a fireball/blastwave model agrees with observations quite well.

However, we notice that afterglows from GRB 970228 and GRB 970508 
have followed simple power-law decays for as long as 190 days and 
80 days respectively, while in Eq.(1), even $t = 30$ d will lead 
to $\gamma \sim 1$. We stress that the overall evolution of the 
postburst fireball can not be regarded as a simple one-phase process. 
One should be careful in applying those scaling laws at later times. 
In fact, it is clear from Eq.(1) that the expansion will become mildly 
relativistic ($2 \leq \gamma \leq 5$) when $t \geq 10$ d, and will 
cease to be relativistic about 30 days later at the latest. 

\section{Our Model} 

We now propose a refined model to describe the full evolution of the 
postburst fireballs. As usual, we suppose that the main GRB occurs at,  
\begin{equation}
   R_0 = ( \frac {3 E_0}{4 \pi n m_{\rm p} c^2 \eta^2} )^{1/3} 
       = 10^{16} E_{51}^{1/3} n_0^{-1/3} \eta_{300}^{-2/3} {\rm cm}, 
\end{equation}
where $\eta = 300 \eta_{300} = E_0/(M_0 c^2)$, $M_0$ is the mass of the 
contaminating baryons. In the subsequent expansion, jump 
conditions for the shock can be described as (Blandford \& McKee 1976): 
\begin{equation}
    n' = \frac{\hat{\gamma} \gamma + 1}{\hat{\gamma} - 1} n,
\end{equation}
\begin{equation}
e'=\frac{\hat{\gamma} \gamma +1}{\hat{\gamma} -1} (\gamma -1) n m_{\rm p} c^2,
\end{equation}
\begin{equation}
    \Gamma ^2 = \frac{(\gamma + 1)[ \hat{\gamma}(\gamma - 1) + 1 ]^2}
    { \hat{\gamma} (2 - \hat{\gamma}) (\gamma - 1) + 2},
\end{equation}
where $n'$ and $e'$ are the electron number density and energy density
\begin{figure}
\begin{picture}(100,160)
\put(0,0){\includegraphics{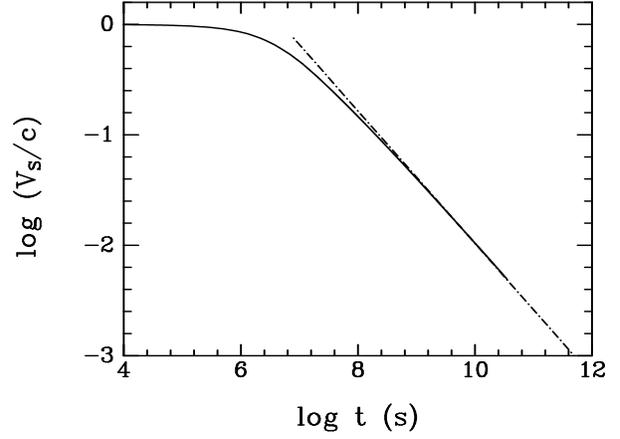}}
\end{picture}
\caption
{ Evolution of the shock's velocity. Full line is our result and dash-dotted 
line is analytic solution for non-relativistic shock wave (Eq.(9))
}
\end{figure}
\begin{figure}
\begin{picture}(100,160)
\put(0,0){\includegraphics{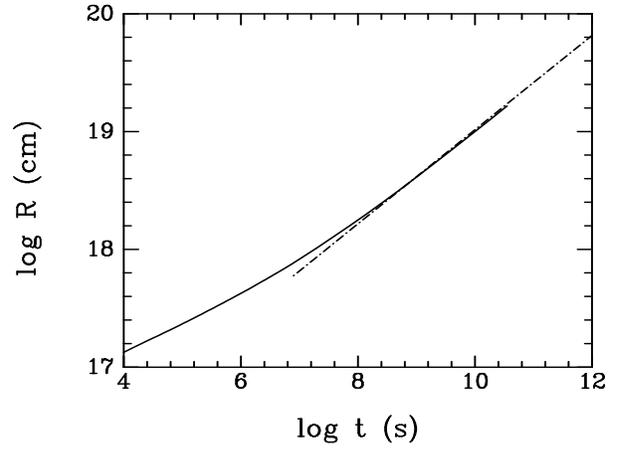}}
\end{picture}
\caption
{ The fireball's radius vs. time. Full line is our result and dash-dotted 
line is analytic solution (Eq.(8)) }
\end{figure}
of the shocked ISM respectively in the frame co-moving with the shell, 
$\Gamma$ is the Lorentz factor of the shock, and $\hat{\gamma}$ is the 
adiabatic index of the ISM, for which we have 
derived an approximation, $\hat{\gamma} \approx (4 \gamma + 1)/(3 \gamma)$, 
consistent with the requirement that $\hat{\gamma} \approx 4/3$ for an 
extremely relativistic blastwave and $\hat{\gamma} \approx 5/3$ for the
non-relativistic Sedov shock. For $\gamma \gg 1$, Eqs.$(3)-(5)$ 
reduce to $n' = 4 \gamma n$,  $e' = 4 \gamma^2 n m_{\rm p} c^2$ and
$\Gamma = \sqrt{2} \gamma$, which are just the starting point of previous
studies. Here we expect that our equations are valid for describing 
relativistic shocks as well as non-relativistic blastwaves.

The kinetic energy of the shocked ISM in the fireball is 
$E_{\rm k} = \sigma \beta^2 \Gamma^2 (4/3) \pi R^3 n m_{\rm p} c^2$ 
(Blandford \& McKee 1976), where 
$\beta = (1 - 1/ \Gamma^2)^{1/2}$, and $\sigma$ is a coefficient: 
$\sigma \rightarrow 0.35 $ when $\beta \rightarrow 1$ and 
$\sigma \rightarrow 0.73 $ when $\beta \rightarrow 0$. 
We will use an approximate equation 
for $\sigma$: $\sigma = 0.73 - 0.38 \beta$. As usual, the expansion 
\begin{figure}
\begin{picture}(100,320)
\put(0,0){\includegraphics{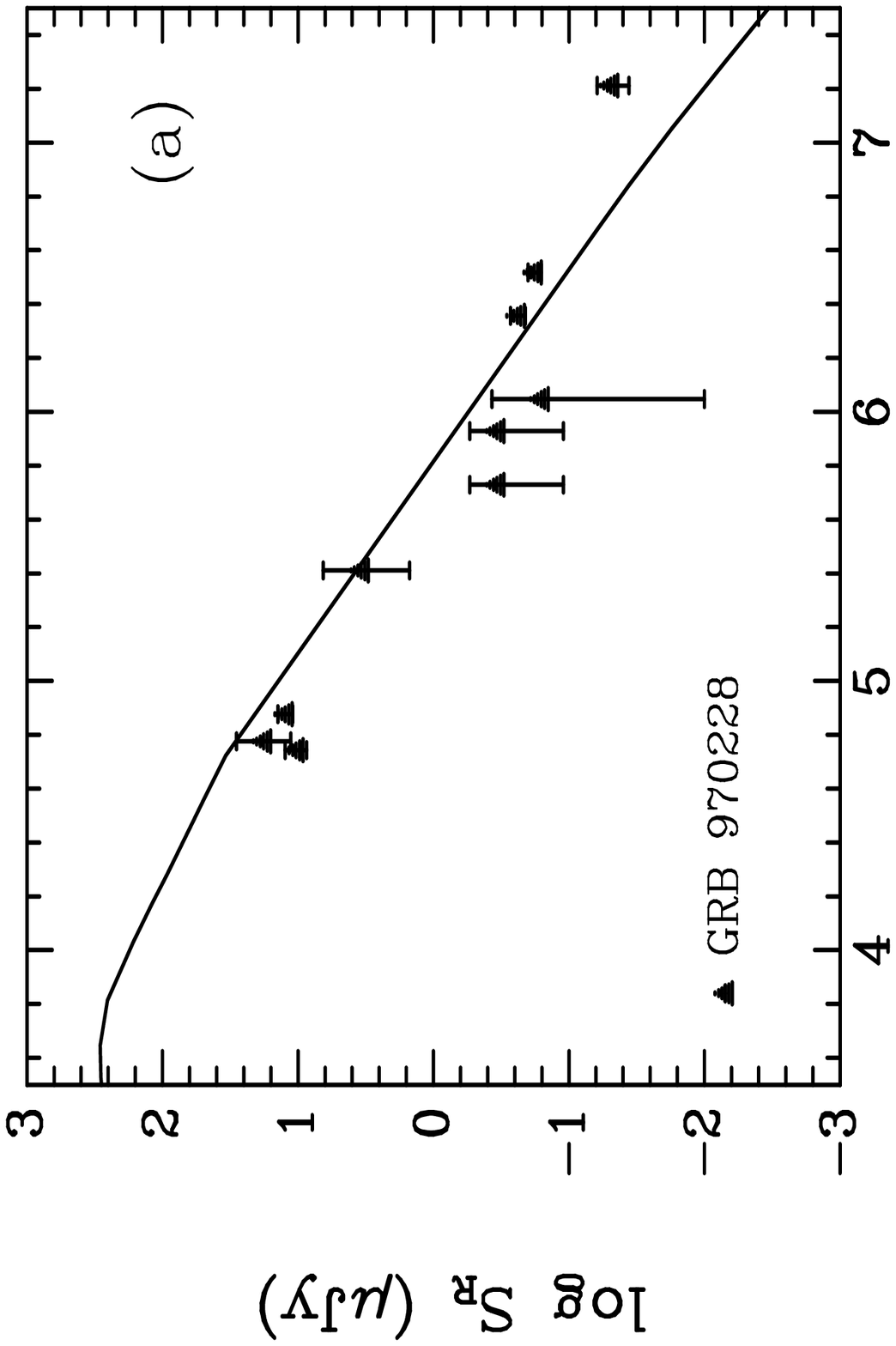}}
\put(0,0){\includegraphics{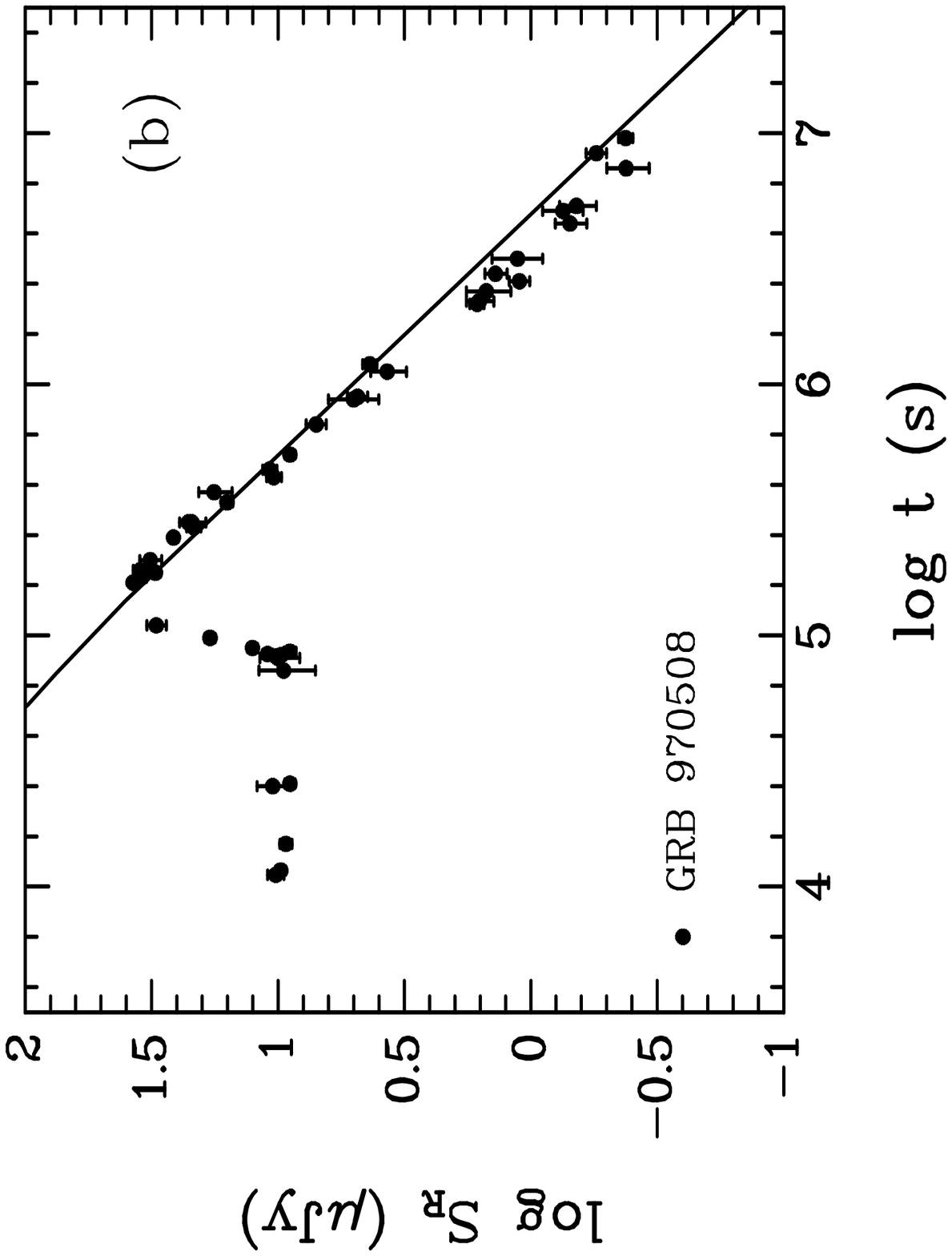}}
\end{picture}
\caption
{ Optical afterglows in R band: {\bf a} GRB 970228, {\bf b} GRB 970508. 
Full lines are our theoretical results. Observed fluxes are from Pedersen 
et al.(1998), Galama et al.(1998a), Garcia et al.(1998) }
\end{figure}
is assumed to be adiabatic, during which energy is conserved, so we have,
\begin{equation}
    \frac {4}{3} \pi \sigma \beta^2 \Gamma^2 R^3 n m_{\rm p} c^2  
        = E_{\rm k} = {{E_0} \over {2}}.
\end{equation}
In order to study the evolution of $\gamma$, we should add the
following differential equation (Huang et al. 1998),  
\begin{equation}
   \frac{dR}{dt} = {\sqrt {\Gamma^2 - 1} \over \Gamma}  
                    \frac {\gamma}{\gamma - \sqrt {\gamma^2 - 1}} c,
\end{equation}

Eqs.(2) $-$ (7) present a perfect description of
the shock. Given initial values of $E_0$ and $M_0$,
$R(t)$ and $\gamma (t)$ can be evaluated numerically.
But under the assumption that $\gamma \gg 1$,
we can derive a simple analytic solution,
$R^4 \approx R_0^4 + 8 k t$, $\gamma \approx (cR^3/k)^{-1/2}$,
where $k=E_0/(4 \pi n m_{\rm p} c)$.  Additionally 
if $R \gg R_0$, that is $t \gg \tau$, where $\tau$ refers to the 
duration of the main GRB, then 
$ R \approx 1.06 \times 10^{16}(E_{51}/n_0)^{1/4} t^{1/4}$ cm,
$ \gamma \approx 273 (E_{51}/n_0)^{1/8} t^{-3/8}$.
These expressions are consistent with previous studies.

\section{Numerical results} 

We have evaluated the propagation of the blastwave 
\begin{figure}
\begin{picture}(100,160)
\put(0,0){\includegraphics{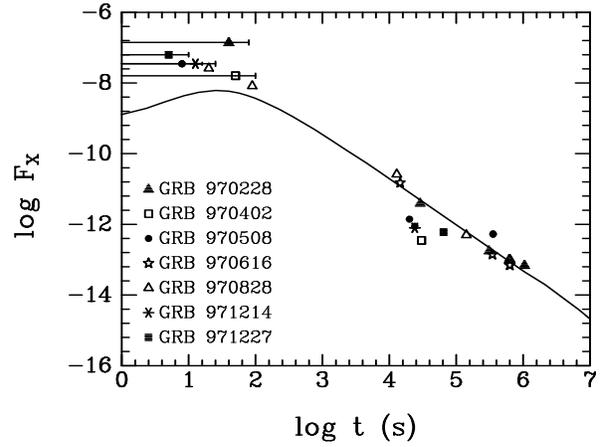}}
\end{picture}
\caption
{ X-ray afterglows ($0.1 - 10$ keV, in unit of erg cm$^{-2}$ s$^{-1}$). 
Full line is our theoretical result. 
For the observed fluxes, please see Dai et al.(1998) }
\end{figure}
numerically, taking 
$E_0 = 10^{51}$ erg, $n=1$ cm$^{-3}$, and $M_0 = 2 \times 10^{-6}$ 
M$_{\odot}$. Fig. 1 is the evolution of $\gamma$, which follows the 
power-law expression of Eq.(1) quite well when 
$10^1 {\rm s } \leq t \leq 10^6 {\rm s}$. However, it is clear that
$\gamma$ ceases to be much larger than 1 when $t \geq 10^6$ s, so that 
Eq.(1) is no longer applicable. 
At least, it is problematic 
to assume that the shock wave was still ultra-relativistic, therefore 
one should be cautious in applying those scaling laws such as 
$R \propto t^{1/4}$, $\gamma \propto t^{-3/8}$, and
$S_{\nu} \propto t^{3(1-p)/4}$. This fact has been 
completely ignored in previous studies.

Fig. 2 illustrates the evolution of the shock wave's 
velocity ($V_{\rm s}$). $V_{\rm s}$ differs markedly from $c$ after 
$\sim 10^6$ s, and the evolution enters the non-relativistic 
phase, for which a simple analytic solution is available (Lang 1980): 
\begin{equation}
    R = 1.15 (E_{\rm k} t^2 / \rho)^{1/5},
\end{equation}
\begin{equation} 
    V_{\rm s} = 3 R / (10 t), 
\end{equation}
where $\rho = n m_{\rm p}$. The dash-dotted line in Fig. 2 is plotted 
according to Eq.(9). Our result is in good agreement with it. 
Fig. 3 shows our evolution of radius, also plotted are results from 
Eq.(8). We see from Figs. 3 and 4 that our refined model 
is applicable for both relativistic and non-relativistic expansion. 

To compare with observations, we need to calculate synchrotron 
radiation from the shocked ISM. As usual, electrons 
in the shocked ISM are assumed to follow a power-law distribution, 
with an index of $p$; the magnetic 
field energy density in the co-moving frame is supposed to be a 
fraction $\xi_{\rm B}^2$ of the energy density,
$B'^2/8 \pi = \xi_{\rm B}^2 e'$; and the electrons are supposed to carry
a fraction $\xi_{\rm e}$ of the energy, 
$\gamma_{\rm e,min} m_{\rm e} c^2 = \xi_{\rm e} \gamma m_{\rm p} c^2$. 
The luminosity distance to the GRB source is designated as $D_{\rm L}$.
We have plotted in Figs. 4 and 5 the calculated R band flux 
densities ($S_R$) and $0.1 - 10$ keV fluxes ($F_X$) respectively, 
and compared them with observed afterglows.

In Fig. 4a, we take $p = 2.5$, $\xi_{\rm B}^2 = 0.01$, 
$\xi_{\rm e} = 1$, and $D_{\rm L} = 3$ Gpc. Other parameters such 
as $E_0$, $n$, $M_0$ are the same as in Fig. 1. 
Generally the theoretical light curve fits 
GRB 970228 well. However, a problem appears at later 
times ($t \geq 10^7$ s) when the blastwave becomes 
non-relativistic. Our model predicts a sharper decline,  
consistent with the conclusion that has been drawn by 
Wijers et al. (1997), but the observed flux decays obviously slower. 
An overall power-law decline lasting for more than $10^7$ s would
require a much larger $E_0$ or a much smaller $n$ to ensure that 
the blastwave is uniformly ultra-relativistic. However, due to 
the limited observational data points, it is arbitrary to make 
any further conclusions.

For the optical afterglow from GRB 970508, a set of physical parameters 
have recently been carefully derived 
as: $E_0 = 3.7 \times 10^{52}$ erg, $n = 0.035$ cm$^{-3}$, 
$\xi_{\rm e} = 0.13$, $\xi_{\rm B}^2 = 0.068$, $p = 2.2$, 
$D_{\rm L} \approx 4$ Gpc (Wijers \& Galama 1998; Galama et al. 1998c,d).  
In Fig. 4b, we have taken these values, with 
$M_0 = 7.4 \times 10^{-5}$ M$_{\odot}$. Since $E_0$ is large and 
$n$ is small, the period during which the ultra-relativistic 
approximation is valid is substantially extended. In fact, 
$\gamma$ keeps to be larger than 1.5 even when $t > 1.2 \times 10^7$ s, 
and the theoretical light curve are well presented by a single 
straight line, which fits observations satisfactorily. We note that
if $p = 2.3$ were assumed, the observational data could be reproduced 
even better. The afterglow from GRB 970508 seemed to decay more 
slowly after about 80 days, implying  
the presence of a constant component (Pedersen et al. 1998; 
Galama et al. 1998a; Garcia et al. 1998), maybe its host galaxy.
The observed optical flux peaked about two days later, possibly  
associated with an X-ray outburst (Piro et al. 1998). This feature 
could not be explained by a simple blastwave model. 
Our theoretical light curve peaks several hours later at optical 
wavelengths and several tens of seconds later in X-rays.

The theoretical X-ray light curve is plotted 
and compared with observations 
in Fig. 5, where parameters are evaluated the same as in Fig. 4a. X-ray 
counterparts usually fade away in less than two weeks when the
expansion is still relativistic, so observed light curves are well 
represented by power-law relations.

\section{Discussion} 

The simple fireball model predicts a power-law decay for GRB afterglows, 
in rough agreement with observations. However, the fact that the
blastwave might enter the non-relativistic phase when the 
afterglows are still detectable has been widely ignored. Here we 
stress that the blastwave generated by a typical fireball will cease 
to be ultra-relativistic in $\sim 10^6$ s. One should be cautious
in applying those simple scaling laws such 
as $\gamma \propto t ^{-3/8}$, $R \propto t^{1/4}$, 
and $S_{\nu} \propto t^{3(1-p)/4}$ (for adiabatic expansion) 
at such late times as $t \geq 10^6$ s.

We have derived refined equations to describe the full evolution 
of postburst fireballs. Our model is consistent with previous 
studies in both the ultra-relativistic phase and the non-relativistic 
phase. The predicted afterglows decay slightly sharper at the 
non-relativistic stage. The observed overall power-law decay lasting 
for several months is usually considered as a strong proof to the 
popular fireball model. Here we point out that it has in fact raised a 
problem, since it requires a large 
$E_0$ and/or a low $n$. For example, recent detection of an enormous 
burst GRB 971214 with $E_0 \sim 3 \times 10^{53}$ erg might  
ameliorate the ultra-relativistic assumption to some 
extent (Kulkarni et al. 1998; Wijers 1998).
The solution could also be that  
there is a persistent energy source at the center of 
the GRB source (Dai \& Lu 1998b).

\acknowledgements 

This work was supported by the National Natural Science Foundation 
and the Foundation of the Ministry of Education of China.

%

\begin{thebibliography}{}

\bibitem[]{}
Blandford R.D., McKee C.F., 1976, Phys. of Fluids 19, 1130

\bibitem[]{}
Dai Z.G., Huang Y.F., Lu T., 1998, ApJ submitted, 
    also see preprint: astro-ph/9806334

\bibitem[]{}
Dai Z.G., Lu T., 1998a, MNRAS, in press, 
    also see preprint: astro-ph/9806305 

\bibitem[]{}
Dai Z.G., Lu T., 1998b, A\&A, 333, L87 

\bibitem[]{}
Fenimore E.E., Madras C.D., Nayakskin S., 1996, ApJ 473, 998

\bibitem[]{}
Fishman G.J., Meegan C.A., 1995, ARA\&A 33, 415

\bibitem[]{}
Galama T.J., Groot P.J., van Paradijs J., et al., 1998a, ApJ, 497, L13 

\bibitem[]{}
Galama T.J., Vreeswijk P.M., Pian E., Frontera~F., Doublier~V., 
   Gonzalez~J.-F., 1998b, IAU Circ. 6895 

\bibitem[]{}
Galama T.J., Wijers R., Bremer M., et al., 1998c, ApJ, 500, L101  

\bibitem[]{}
Galama T.J., Wijers R., Bremer M., et al., 1998d, ApJ, 500, L97  

\bibitem[]{}
Garcia M.R., Callanan P.J., Moraru D., et al., 1998, ApJ, 500, L105  

\bibitem[]{}
Huang Y.F., Dai Z.G., Wei D.M., Lu T., 1998, MNRAS, in press, 
    also see preprint: astro-ph/9806333 

\bibitem[]{}
Klebesadel R.W., Strong I.B., Olson R.A., 1973, ApJ 182, L85

\bibitem[]{}
Kulkarni S.R., Djorgovski S.G., Ramaprakash A.N., et al., 1998, Nat 393, 35

\bibitem[]{}
Lang K.R., 1980, In: Astrophysical Formulae. Springer, Berlin, p. 302

\bibitem[]{}
M\'{e}sz\'{a}ros P., Rees M.J., 1997, ApJ 476, 232 

\bibitem[]{}
M\'{e}sz\'{a}ros P., Rees M.J., Papathanassiou H., 1994, ApJ 432, 181

\bibitem[]{}
Pedersen H., Jaunsen A.O., Grav T., et al., 1998, ApJ 496, 311

\bibitem[]{}
Piro L., Amati L., Antonelli L.A., 1998, A\&A 331, L41

\bibitem[]{}
Sari R., 1997, ApJ 489, L37

\bibitem[]{}
Tavani M., 1997, ApJ 483, L87

\bibitem[]{}
Waxman E., 1997a, ApJ 485, L5

\bibitem[]{}
Waxman E., 1997b, ApJ 489, L33

\bibitem[]{}
Wijers R., 1998, Nat 393, 13

\bibitem[]{}
Wijers R., Galama T.J., 1998, ApJ, submitted, also see 
   preprint: astro-ph/9805341

\bibitem[]{}
Wijers R., Rees M.J., M\'{e}sz\'{a}ros P., 1997, MNRAS 288, L51



\end{thebibliography}
\end{document}